\voffset=0.3cm\input boxedeps.tex\input epsf.tex\SetRokickiEPSFSpecial
\HideDisplacementBoxes
\hfuzz=5pt


\font \tensans                = cmss10
\font \fivesans               = cmss10 at 5pt
\font \sixsans                = cmss10 at 6pt
\font \sevensans              = cmss10 at 7pt
\font \ninesans               = cmss10 at 9pt
\newfam\sansfam
 \def\sans{\fam\sansfam\tensans}%
 \textfont\sansfam=\tensans \scriptfont\sansfam=\sevensans
 \scriptscriptfont\sansfam=\sixsans



\def\=def{\buildrel\rm def\over =}

\font\manual=cmssbx10 at 24pt
\def\dbend{\hbox{\manual Z\quad}}
\def\d@nger{\medbreak\begingroup\clubpenalty=10000\parshape 3 1truecm 13truecm
1truecm 13truecm 0cm \hsize
\noindent
\vbox to5pt{\hbox to5pt{\hss\dbend\hfil}\vss}\kern -5pt}
\outer\def\danger #1\par{\d@nger #1\endgraf\endgroup\medbreak}


\def\bulet{{\petit $\bullet$\quad}}

\def\N{{\rm I\!N}}

\def\T{{\rm I\kern-4.3pt T}}

\def\C{{\mathchoice {\setbox0=\hbox{$\displaystyle\rm C$}\hbox{\hbox
to0pt{\kern0.4\wd0\vrule height0.96\ht0\hss}\box0}}
{\setbox0=\hbox{$\textstyle\rm C$}\hbox{\hbox
to0pt{\kern0.4\wd0\vrule height0.96\ht0\hss}\box0}}
{\setbox0=\hbox{$\scriptstyle\rm C$}\hbox{\hbox
to0pt{\kern0.4\wd0\vrule height0.96\ht0\hss}\box0}}
{\setbox0=\hbox{$\scriptscriptstyle\rm C$}\hbox{\hbox
to0pt{\kern0.4\wd0\vrule
height0.96\ht0\hss}\box0}}}}

\def\Q{{\mathchoice {\setbox0=\hbox{$\displaystyle\rm
Q$}\hbox{\raise 0.15\ht0\hbox to0pt{\kern0.4\wd0\vrule
height0.8\ht0\hss}\box0}} {\setbox0=\hbox{$\textstyle\rm
Q$}\hbox{\raise 0.15\ht0\hbox to0pt{\kern0.4\wd0\vrule
height0.8\ht0\hss}\box0}} {\setbox0=\hbox{$\scriptstyle\rm
Q$}\hbox{\raise 0.15\ht0\hbox to0pt{\kern0.4\wd0\vrule
height0.7\ht0\hss}\box0}}
{\setbox0=\hbox{$\scriptscriptstyle\rm Q$}\hbox{\raise
0.15\ht0\hbox to0pt{\kern0.4\wd0\vrule height0.7\ht0\hss}
\box0}}}}

\def\Z{{\mathchoice {\hbox{$\sans\textstyle Z\kern-0.4em Z$}}
{\hbox{$\sans\textstyle Z\kern-0.4em Z$}}
{\hbox{$\sans\scriptstyle Z\kern-0.3em Z$}}
{\hbox{$\sans\scriptscriptstyle Z\kern-0.2em Z$}}}}




\def\defnomcs#1{\expandafter\def\csname #1\endcsname}
\def\nomcs#1{\csname #1\endcsname}

\def\warning#1{\immediate\write16{#1}\immediate\write16{}}

\def\ifnextchar#1#2#3{\let\tempora=#1\def\temporb{#2}%
  \def\temporc{#3}\futurelet\tempord\ifnextcharchoix}
\def\ifnextcharchoix{\ifx\tempord\tempora\let\temporf=\temporb%
                     \else\let\temporf=\temporc\fi\temporf}

\long\def\ifundefined#1#2#3{\expandafter%
               \ifx\csname #1\endcsname\relax#2\else#3\fi}

\def\vide{}
\def\for#1:=#2!do#3{\edef\tempa{#2}\ifx\tempa\vide{}\else%
\edef\tempb{#2,!}\expandafter\bouclefor\tempb@#1#3@\fi}
\def\bouclefor#1,#2@#3#4@{\def#3{#1}#4\ifx#2!{}\else%
                                 \expandafter\bouclefor#2@#3#4@\fi}

\newwrite\toc
\immediate\openout\toc=\jobname.to
\immediate\write\toc{\string\relax}
\newwrite\toa
\immediate\openout\toa=\jobname.toa
\immediate\write\toa{\string\relax}


\newwrite\index
\immediate\openout\index=\jobname.indx
\immediate\write\index{\string\relax}
\def\ind#1{\write\index {#1, \the\pageno\string\hfill\string\break}}


\newwrite\refce
\immediate\openout\refce=\jobname.ref
\immediate\write\refce{\string\relax}

\def\ref#1{\ifundefined{ref#1}{??
\warning{  Attention, la reference "#1", page \the\pageno,
section \the\secno.\the\subsecno,
est inconnue}}{\hbox{\nomcs{ref#1}}}}

\def\refno#1{\ifundefined{refno#1}{??
\warning{  Attention, la reference "#1", page \the\pageno,
section \the\secno.\the\subsecno,
est inconnue}}{\hbox{\nomcs{refno#1}}}}


 \newwrite\sol
 \immediate\openout\sol=\jobname.sol
 \immediate\write\sol{\string\relax}

     
     \newcount\secno
     \newcount\subsecno
     \newcount\thmno
     \newcount\exno
     \newcount\figno
					\newcount\formuleno


\outer\def\partie{\normalpartie}

\def\normalpartie #1\par{
 \secno=0 \thmno=0 \exno=0 \figno=0
\formepartie{#1}
}

\font\fontpart=cmbx10 at 12pt

\def\formepartie#1{\vfill\eject
\vglue1.5cm
\centerline{\fontpart #1}
\xdef\rightheadline{\ignorespaces{\fonttitrecourant #1}\unskip\hfil\llap{\noexpand\folio}}
\xdef\leftheadline{\rlap{\noexpand\folio}\hfil\ignorespaces{\fonttitrecourant
 #1}\unskip}
\nobreak\vskip1cm\noindent}

\def\titgauche #1\par{
\xdef\leftheadline{\rlap{\noexpand\folio}\hfil\ignorespaces{\fonttitrecourant #1}\unskip}}


\outer\def\section{\ifnextchar[{\labelsection}%
  {\ifnextchar*{\etoilesection}{\normalsection}}}

\def\etoilesection* #1\par{
  \formesectionetoile{#1}}

\def\normalsection #1\par{\advance\secno by1 \subsecno=0
  \formesection{#1}}

\def\labelsection[#1] #2\par{\advance\secno by1 \subsecno=0
  \immediate\write\refce{\string\defnomcs{ref#1}{\hbox{Section 
  \the\secno}}}
\immediate\write\refce{\string\defnomcs{refno#1}{\the\secno}}
  \formesection{#2}
}

\font\fontsection=cmbx10 at 12pt

\def\formesectionetoile#1{
  \vskip0pt plus.07\vsize\penalty-300
  \vskip0pt plus-.07\vsize
  \bigskip\vskip\parskip
  \medbreak\noindent
  {\fontsection #1}\par\vskip\parskip
  \nobreak\medskip}

\def\formesection#1{
  \vskip0pt plus.075\vsize\penalty-1050
  \vskip0pt plus-.075\vsize\mark{\the\secno\enspace#1}
  \bigskip\vskip\parskip
  \medbreak\noindent
  {\fontsection \the\secno\enspace#1}\par
  \nobreak\medskip}


\outer\def\subsection{\ifnextchar[{\labelsubsection}%
  {\ifnextchar*{\etoilesubsection}{\normalsubsection}}}

\def\etoilesubsection* #1\par{
  \formesubsectionetoile{#1}}

\def\normalsubsection #1\par{\advance\subsecno by1
  \formesubsection{#1}}

\def\labelsubsection[#1] #2\par{\advance\subsecno by1
  \immediate\write\refce{\string\defnomcs{ref#1}{\hbox{Section
  \the\secno.\the\subsecno}}}
\immediate\write\refce{\string\defnomcs{refno#1}{\the\secno.\the\subsecno}}
  \formesubsection{#2}}

\font\fontsubsection=cmbx10

\def\formesubsectionetoile#1{
  \vskip0pt plus.032\vsize\penalty-320
  \vskip0pt plus-.032\vsize
  \vskip\parskip\medbreak\noindent
  {\fontsubsection #1}\par
  \nobreak\medskip}

\def\formesubsection#1{
  \vskip0pt plus.043\vsize\penalty-630
  \vskip0pt plus-.043\vsize
  \medbreak\noindent
  {\fontsubsection\the\secno.\the\subsecno\enspace#1}\par
  \nobreak\medskip
  {\let\folio=0\edef\next{\write\toc{\string\the\subsecno;#1;\folio\par}}\next}
}



\long\def\normalenonce#1#2\fin\par{\advance\thmno by1
  \formeenonce{#1}{#2}}

\long\def\labelenonce#1[#2]#3\fin\par{\advance\thmno by1
  \immediate\write\refce{\string\defnomcs{ref#2}{\hbox{\lowercase{#1}
  \the\thmno}}}
\immediate\write\refce{\string\defnomcs{refno#2}{\the\thmno}}
  \formeenonce{#1}{#3}}

\def\defi{\def\titre{\ifanglais{Definition}\else{D\'efinition}\fi}
  \ifnextchar[{\labelenonce\titre}{\normalenonce\titre}}

\def\thm{\def\titre{\ifanglais{Theorem}\else{Th\'eor\`eme}\fi}
  \ifnextchar[{\labelenonce\titre}{\normalenonce\titre}}

\def\lemme{\def\titre{\ifanglais{Lemma}\else{Lemme}\fi}
  \ifnextchar[{\labelenonce\titre}{\normalenonce\titre}}

\def\corol{\def\titre{\ifanglais{Corollary}\else{Corollaire}\fi}
  \ifnextchar[{\labelenonce\titre}{\normalenonce\titre}}


\font\typeenonce=cmbx10
\font\fontenonce=cmsl10

\long\def\formeenonce#1#2{\medbreak\noindent{\typeenonce#1
  \the\thmno\enspace}
  {\fontenonce #2}\par
  \ifdim\lastskip<\medskipamount
  \removelastskip\penalty55\medskip\fi}


\long\def\enoncesimple#1#2\fin\par{\advance\thmno by1
  \formeenoncesimple{#1}{#2}}
 
\long\def\labelenoncesimple#1[#2]#3\fin\par{\advance\thmno by1
  \immediate\write\refce{\string\defnomcs{ref#2}{\hbox{\lowercase{#1} 
  \the\thmno}}}
\immediate\write\refce{\string\defnomcs{refno#2}{\the\thmno}}
  \formeenoncesimple{#1}{#3}}

\def\ex{\def\titre{\ifanglais{Example}\else{Exemple}\fi}
  \ifnextchar[{\labelenoncesimple\titre}{\enoncesimple\titre}}

\def\remarque{\def\titre{\ifanglais{Remark}\else{Remarque}\fi}
  \ifnextchar[{\labelenoncesimple\titre}{\enoncesimple\titre}}


\font\typeenoncesimple=cmcsc10
\font\fontenoncesimple=cmr10

\long\def\formeenoncesimple#1#2{\medbreak\noindent{\typeenoncesimple#1 
  \the\thmno\enspace}
  {\fontenoncesimple #2}\par
  \ifdim\lastskip<\medskipamount
  \removelastskip\penalty55\medskip\fi}


\font\fonttitrepreuve=cmsl10

\def\preuve{\noindent{\fonttitrepreuve
\ifanglais{Proof}\else{D\'emonstration}\fi.\enspace}}

\def\QED{\hbox{\rlap{$\sqcap$}$\sqcup$}}

\def\qed{\ifmmode\eqno\QED\else{\parfillskip=0pt \enspace\hfill
  \rlap{$\sqcap$}$\sqcup$ \ifdim\lastskip<\medskipamount
  \removelastskip\penalty55\medskip\fi \parfillskip=0pt
  plus1fil}\fi}


  %
  %
  %
  %

\font \sixbf                  = cmbx6

\font \ninebf                 = cmbx9

\font \sixi                   = cmmi6
\font \ninei                  = cmmi9

\font \tbmss                  = cmmib10 scaled 600

\font \tbms                   = cmmib10 scaled 833

\font \sixrm                  = cmr6
\font \ninerm                 = cmr9
\font \ninesl                 = cmsl9
\font \tensans                = cmss10
\font \fivesans               = cmss10 at 5pt
\font \sixsans                = cmss10 at 6pt
\font \sevensans              = cmss10 at 7pt
\font \ninesans               = cmss10 at 9pt

\font \sixsy                  = cmsy6

\font \ninesy                 = cmsy9

\font \nineit                 = cmti9
\font \ninett                 = cmtt9
\def\petit{\def\rm{\fam0\ninerm}%
\textfont0=\ninerm \scriptfont0=\sixrm \scriptscriptfont0=\fiverm
 \textfont1=\ninei \scriptfont1=\sixi \scriptscriptfont1=\fivei
 \textfont2=\ninesy \scriptfont2=\sixsy \scriptscriptfont2=\fivesy
 \def\it{\fam\itfam\nineit}%
 \textfont\itfam=\nineit
 \def\sl{\fam\slfam\ninesl}%
 \textfont\slfam=\ninesl
 \def\bf{\fam\bffam\ninebf}%
 \textfont\bffam=\ninebf \scriptfont\bffam=\sixbf
 \scriptscriptfont\bffam=\fivebf
 \def\sans{\fam\sansfam\ninesans}%
 \textfont\sansfam=\ninesans \scriptfont\sansfam=\sixsans
 \scriptscriptfont\sansfam=\fivesans
 \def\tt{\fam\ttfam\ninett}%
 \textfont\ttfam=\ninett
 \normalbaselineskip=11pt
 \parskip=0pt
 \setbox\strutbox=\hbox{\vrule height7pt depth2pt width0pt}%
 \normalbaselines\rm
\def\vec##1{{\textfont1=\tbms\scriptfont1=\tbmss
\textfont0=\ninebf\scriptfont0=\sixbf
\mathchoice{\hbox{$\displaystyle##1$}}{\hbox{$\textstyle##1$}}
{\hbox{$\scriptstyle##1$}}{\hbox{$\scriptscriptstyle##1$}}}}}

\long\def\enonceexo#1#2\fin\par{\advance\exno by1
  \formeenonceexo{#1}{#2}}
 
\long\def\labelenonceexo#1[#2]#3\fin\par{\advance\exno by1
  \immediate\write\refce{\string\defnomcs{ref#2}{\hbox{\lowercase{#1}
  \the\exno}}}
\immediate\write\refce{\string\defnomcs{refno#2}{\the\exno}}
  \formeenonceexo{#1}{#3}}


\font\typeenonceexo=cmcsc10 at 9pt

\def\diam{\parfillskip=0pt \enspace\hfill
  $\diamondsuit$ \ifdim\lastskip<\medskipamount
  \removelastskip\penalty55\medskip\fi \parfillskip=0pt
  plus1fil}

\long\def\formeenonceexo#1#2{\medbreak\noindent
{\typeenonceexo#1 
  \the\exno\enspace}
  {\petit\baselineskip=12.6truept
 #2\diam}\par
  \ifdim\lastskip<\medskipamount
  \removelastskip\penalty55\medskip\fi}



 \chardef\other=12

 \long\def\solution{
   \immediate\write\sol{}
   \immediate\write\sol{{\string\par \string\medbreak \string\bf
   \string\the\exno.}}
   \copytoblankline}

 \def\copytoblankline{\begingroup\setupcopy\copysol}

 \def\setupcopy{\def\do##1{\catcode`##1=\other}\dospecials
   \catcode`\|=\other\catcode`\:=12\catcode`\;=12
    \catcode`\!=12\catcode`\?=12
\obeylines}

 { \catcode`\/=0 \catcode`\\=\other
   /gdef/finsol{\fin}
 } 

 {\obeylines \gdef\copysol#1
   {\def\next{#1}%
   \ifx\next\finsol\let\next=\endgroup %
   \else\immediate\write\sol{\next} \let\next=\copysol\fi\next}}

%
%

\def\program{\baselineskip=12.5truept\font\motsCles=cmtcsc10
\tt \cleartabs
\def\esp##1{\cleartabs \dimen0=.5cm \multiply\dimen0 by ##1 \kern\dimen0}
\def\<##1>{\hbox{\motsCles##1\/}}
}


\newwrite\biblio     \newcount\bibno


\def\bibi[#1]{\advance\bibno by1
\immediate\write\refce{\string\defnomcs{refbib#1}{\the\bibno}}
  \item{[\the\bibno]}}

\def\cit#1{\hbox{\nomcs{refbib#1}}}


\def\references{\vskip0pt plus.3\vsize \penalty-250
  \vskip0pt plus-.3\vsize \bigskip \vskip \parskip
\leftline{\fontreftitre
  \ifanglais{References}\else{R\'{e}f\'{e}rences}\fi}
  \nobreak\bigskip \bgroup\fontref}
\let\finreferences=\egroup

\def\normalbib#1{\advance\bibno by1
\immediate\write\biblio{\string\defnomcs{refbib#1}{\the\bibno}}
  \item{[\the\bibno]}}

\def\labelbib[#1]#2{\advance\bibno by 1
\immediate\write\biblio{\string\defnomcs{refbib#2}{#1}}
  \par\hang\noindent[#1]\enspace}

\def\cite#1{[\precite{#1}]}
\def\precite#1{\def\virgule{}\for\opcit:=#1!do%
  {\virgule\def\virgule{,}%
  \ifundefined{refbib\opcit}{??%
  \warning{  Attention, la reference bibliographique "\opcit",
  page \the\pageno, Section \the\secno.\the\subsecno\space est
  inconnue}}\hbox{\nomcs{refbib\opcit}}}}



%
%
%
%
%
%

\def\picture #1 by #2 (#3){
  \vbox to #2{
    \hrule width #1 height 0pt depth 0pt
    \vfill
    \special{picture #3} 
    }
  }

\def\scaledpicture #1 by #2 (#3 scaled #4){{
  \dimen0=#1 \dimen1=#2
  \divide\dimen0 by 1000 \multiply\dimen0 by #4
  \divide\dimen1 by 1000 \multiply\dimen1 by #4
  \picture \dimen0 by \dimen1 (#3 scaled #4)}
  }

\font\fontfig=cmcsc10
\font\fontlegendefig=cmr10
\long\def\enoncefig (#1) #2 by #3 scaled #4 #5{
  \par\medbreak
  \vbox{$$\hbox{\scaledpicture #2 by #3 (#1 scaled #4)}$$
  \centerline{\fontlegendefig #5}}
  \par\ifdim\lastskip<\medskipamount
  \removelastskip\penalty55\medskip\fi}


\long\def\labelenoncefig#1[#2] #3 by #4 scaled #5 #6{\global\advance\figno by1
  \immediate\write\refce{\string\defnomcs{ref#2}{\hbox{#1
  {\the\figno}}}}
  \immediate\write\refce{\string\defnomcs{refno#2}{\the\figno}}
  \par\medbreak
  \vbox{$$\hbox{\scaledpicture #3 by #4 (#2 scaled #5)}$$
  \centerline{{\fontfig#1 
  \kern-.30em.\the\figno}\enspace{\fontlegendefig #6}}}
  \par\ifdim\lastskip<\medskipamount
  \removelastskip\penalty55\medskip\fi}


\long\def\enoncefigeps (#1) #2{
  \par\medbreak
  \centerline{\epsfbox{#1.ps}}
  \centerline{\fontlegendefig #2}
  \par\ifdim\lastskip<\medskipamount
 \removelastskip\penalty55\medskip\fi}


\long\def\labelenoncefigeps#1[#2] scaled #3  #4 {\global\advance\figno by1
  \immediate\write\refce{\string\defnomcs{ref#2}{\hbox{\lowercase{#1}
  {\fontenoncesimple\the\figno}}}}
  \immediate\write\refce{\string\defnomcs{refno#2}{
  \fontenoncesimple\the\figno}}
  \par\medbreak
\begingroup
\def\epsfsize##1##2{#3##1}
\centerline{\epsfbox{#2.ps}}\nobreak\centerline{{\fontfig#1 \the\figno}\enspace{\fontlegendefig #4}}\endgroup
  \par\ifdim\lastskip<\medskipamount
 \removelastskip\penalty55\medskip\fi}

\long\def\labelenoncefigs#1[#2] scaled #3  #4{\global\advance\figno by1
 \immediate\write\refce{\string\defnomcs{ref#2}{\hbox{\lowercase{#1}
  {\the\figno}}}}
  \immediate\write\refce{\string\defnomcs{refno#2}{\the\figno}}
  \par\medbreak
\begingroup
\centerline{\BoxedEPSF{#2.ps scaled #3}}\vglue2pt\nobreak\centerline{\strut{\fontfig#1 \the\figno}\enspace{\fontlegendefig #4}}\endgroup
  \par\ifdim\lastskip<\medskipamount
  \removelastskip\penalty55\medskip\fi}

\def\figs{\def\titre{Figure}
  \ifnextchar[{\labelenoncefigs\titre}{\enoncefigs}}



\def\eqnum{\eqno}


\newtoks\formeformule
\formeformule={(\the\formuleno)}

\font\fontnumeroform=cmr10
\long\def\labelformule[#1]{\global\advance\formuleno by1%
\eqnum\the\formeformule\immediate\write\refce{\string\defnomcs{ref#1}%
{\fontnumeroform (\the\formuleno)}}}
 
\long\def\normaleformule{\global\advance\formuleno by1%
\eqnum\the\formeformule}

\long\def\labelfformule[#1]{\global\advance\formuleno by1%
{(\the\formuleno)}\immediate\write\refce{\string\defnomcs{ref#1}%
{\fontnumeroform (\the\formuleno)}}}


\outer\def\bye{\par\vfill\supereject\end}


\def\item{\dimen10=1cm \par \hskip\dimen10 \hangindent\dimen10 \textindent}

              
\newif\ifanglais
 \anglaistrue

\ifanglais{}
\else
  \catcode`\@=11
  \pretolerance=500 \tolerance=1000 \brokenpenalty=5000
  \catcode`\;=\active
  \def;{\relax\ifhmode\ifdim\lastskip>\z@\unskip\fi\kern.2em\fi\string;}
  \catcode`\:=\active
  \def:{\relax\ifhmode\ifdim\lastskip>\z@\unskip\fi\penalty\@M\ \fi\string:}
  \catcode`\!=\active
  \def!{\relax\ifhmode\ifdim\lastskip>\z@\unskip\fi\kern.2em\fi\string!}
  \catcode`\?=\active
  \def?{\relax\ifhmode\ifdim\lastskip>\z@\unskip\fi\kern.2em\fi\string?}
  \frenchspacing
  \catcode`\@=12\fi


\hsize=14truecm
\vsize=24truecm

\baselineskip=13truept
\parskip=4pt plus 2pt minus 2pt
\parindent=0pt 


\font \fonttitrecourant                 = cmsl9
\newif\ifprempage\prempagefalse
\newtoks\hautpagetitre       \hautpagetitre={\hfil}
\newtoks\baspagetitre        \baspagetitre={\hss\tenrm\folio\hss}

\headline={\petit\ifprempage\the\hautpagetitre\else
\ifodd\pageno
\rightheadline\else\leftheadline\fi\fi}
\footline={\petit\ifprempage\the\baspagetitre\global\prempagefalse\else\nopagenumbers\fi}



\hfuzz=7pt\voffset = -1cm

\normalbaselineskip=12pt
\parskip=2pt plus 1pt minus 1pt
\anglaistrue
\hsize=16truecm
\vsize=24truecm

\prempagetrue
\baselineskip=12pt


\vskip 1cm
\centerline{\bf Eug\'enie Foustoucos\ \ Ir\`ene Guessarian}
\vskip3cm
\centerline{\fontpart Contact Author\ \ Ir\`ene Guessarian}

\vskip 1cm\centerline{Address: LIAFA, Universit\'e Paris 7, case 7014,
    2 Place Jussieu, 75251 Paris Cedex 5, France.}
\vskip0.5cm
\centerline{email: {\bf \ \ ig@liafa.jussieu.fr}}

\vskip2cm
\centerline{Classification of paper:  {\fontpart Logic in Computer Science}}
\prempagetrue

\partie {Complexity of Monadic inf-datalog. Application to Temporal Logic.}

\titgauche { Eug\'enie Foustoucos \ \ Ir\`ene Guessarian}

\centerline{\fontpart Eug\'enie Foustoucos{\rm\footnote*{aflaw@otenet.gr, eugenie@math.uoa.gr, MPLA, Department of Mathematics,
National and \break Capodistrian University of Athens, Panepistimiopolis, 15784
 Athens, Greece.} }
 \ \ Ir\`ene Guessarian{\rm\footnote{**}{{\bf Corresponding author:
ig@liafa.jussieu.fr} LIAFA, UMR 7089, Universit\'e Paris 7, case 7014,
  2 Place Jussieu, 75251 Paris Cedex 5, France.}}}
\vskip0.5cm\noindent
{\petit
{\bf Abstract:} In [\cit{gfaa}] we defined Inf-Datalog and
characterized the fragments of   Monadic inf-Datalog that have the
   same expressive power as Modal Logic (resp. $CTL$, alternation-free 
   Modal $\mu$-calculus and Modal $\mu$-calculus). 
We study here the time and space
complexity of evaluation of Monadic inf-Datalog programs
on finite models.
We deduce a new unified proof that  model checking has

1.  linear data and program complexities (both in time and space)
for $CTL$ and alternation-free   Modal $\mu$-calculus, and

2.  linear-space (data and program)  complexities, 
linear-time program complexity
and polynomial-time data complexity 
for $L\mu_k$ (Modal $\mu$-calculus with fixed alternation-depth at most $k$).}
 
\section Introduction

The  model checking problem for a  logic $\cal A$
consists in verifying whether a  formula $\phi$ of $\cal A$
is satisfied in a given structure $\cal K$. In computer-aided verification,
 $\cal A$ is a temporal
logic {\it i.e.} a modal logic used for the description of the temporal
ordering of events and  $\cal K$ is a (finite) Kripke
structure {\it i.e.} a graph equipped with a labelling function
associating with each node $s$ the finite set of propositional variables of
$\cal A$ that are true at node $s$.

Our approach to temporal
logic model checking is based on the close relationship
between model checking and Datalog query evaluation:  a Kripke
structure $\cal K$ can be seen as a relational database 
 and a  formula $\phi$ can be thought of
as a Datalog query $\cal Q$.  In this context, the model checking problem for
$\phi$ in $\cal K$ corresponds to the evaluation of $\cal Q$ on
input database $\cal K$.

In [\cit{gfaa}] we introduced the language inf-Datalog, which
extends usual least fixpoint semantics of Datalog with greatest
 fixpoint semantics, we
gave translations from various temporal logics
 (CTL, ETL, alternation-free Modal $\mu$-calculus, and Modal
 $\mu$-calculus,  by increasing order of expressive power [\cit{e90}])
  into Monadic inf-Datalog and we also gave translations
 from fragments of Monadic inf-Datalog into these logics.
In this paper we give upper bounds for evaluating Monadic
inf-Datalog queries: we describe an algorithm evaluating Monadic
inf-Datalog queries and analyze its complexity with respect to
the size of the database ({\it data complexity}) and its
 complexity with respect to the size of the program
({\it program complexity}). The data complexity is polynomial-time  and
 becomes linear when the program is
stratified (with respect to least and greatest
fixed points nesting): from this we derive a unified proof of the
(known) linear-time data complexity of the  model checking problem
for CTL, ETL and the alternation-free $\mu$-calculus.
The program complexity of our algorithm 
is linear-time and linear-space, and the
data complexity is linear-space too.  Using then our translations in
[\cit{gfaa}] between the temporal logic paradigm and the database
paradigm, we can deduce upper bounds for the complexity of
the model checking problem
in the aforementioned temporal logics.  This is worthwhile
especially for the space complexity which is less studied than the
time complexity.

\section Definitions

 The basic definitions about  Datalog can be found in [\cit{gfaa},\cit{ggv}],
and the basic definitions about  the $\mu$-calculus can be found in 
[\cit{AN01},\cit{E}]. We proceed directly with the definition of inf-Datalog.

\defi[infdatalog] 
An {\sl inf-Datalog program} is a Datalog program where
some IDB predicates are tagged with an overline 
indicating that they must be computed as greatest fixed points, and
where in addition, for each set of mutually recursive IDB predicates  including
both tagged and
untagged IDB predicates, the order of evaluation of the IDB predicates in the
set is specified.

    An  inf-Datalog program is said to be {\bf monadic} if all the
predicates occurring in the heads of the rules have arity at most one. 
An  inf-Datalog program is said to be  {\bf stratified} 
if no tagged IDB
predicate is mutually recursive with an untagged IDB
predicate. 
\fin

Our approach  allows us to define  some recursive 
predicates  without initialization rules (non-recursive rules with this
predicate in the head); such recursive predicates must be tagged. This approach
is necessary in order to be able to express properties such as fairness
(something must happen infinitely often).

The above notion of stratification is the natural counterpart (with respect to
greatest fixed points) of the
well-known stratification with respect to negation.
We give an example of a stratified inf-Datalog program.

{\petit
\ex[AGinf1] Consider as database an infinite 
  full binary tree,
  with two EDB predicates $Suc_0$ and $Suc_1$ denoting respectively the first
  successor and the second successor, and a unary EDB predicate $p$
  (which is meant to state some property of the nodes of the tree). 
  The program $P$ below, has as
  IDB predicates $\overline \theta$ (computed as a greatest fixed point) and
  $\varphi$  (computed as a least fixed point)

  $P\colon\left\{\eqalign{
                    {\overline \theta} (x)&\longleftarrow
                    p(x),  Suc_0(x,y),Suc_1(x,z), {\overline \theta} (y),
                        {\overline \theta} (z)\cr
                     \varphi (x)&\longleftarrow {\overline \theta}(x)\cr
                      \varphi (x)&\longleftarrow Suc_0(x,y),Suc_1(x,z), 
                    {\varphi} (y), {\varphi} (z)\cr
                    }
                    \right. $

The IDB predicate $\overline \theta$ in this program implements the modality 
${\bf A}G p$ on the infinite full binary tree, and
the IDB predicate $\varphi$ implements the modality 
${\bf A}F{\bf A}G
p$: ${\bf A}G p$ means that $p$ is always true on all paths,
and  ${\bf A}F{\bf A}G p$ means that, on every
path we will eventually (after a finite number of steps) reach a state
wherefrom $p$ is always true on all paths. 
$G p$ is expressed by the $CTL$ path formula $\bot {\bf \widetilde{U}} p$
and ${\bf A}F{\bf A}Gp$ is expressed by the $CTL$ state formula 
${\bf A}\big(\top {\bf U} {\bf A}(\bot {\bf \widetilde{U}} p)\big)$.
The $\mu$-calculus analog is the $L\mu_1$ expression $\mu \varphi.\Big(\nu
 \theta.(p\wedge {\bf A}\circ\; \theta)\bigvee {\bf A}\circ\; \varphi\Big)$.
\fin

}
\prempagefalse
\section Complexity of Monadic inf-Datalog 

\thm[strat] Let $P$ be a stratified Monadic inf-Datalog program having $I$
IDB symbols, and $D$ a relational database having $n$ elements in its domain,
 then the set of {\bf all} $I$ queries defined by $P$ 
(of the form $(P,\varphi)$, where $\varphi$ is an IDB of $P$)
can be evaluated in
time $n\times I$ and space $n\times I$.
\fin

\preuve By induction on the number $p$ of strata.  Assume $P$ has a single
stratum, and, e.g. all IDBs are untagged, hence computed as least fixed points.
Let $\varphi_1,\ldots,\varphi_I$ be the IDBs, then the answer
$f_1,\ldots,f_I$ to the set of queries 
$(P,\varphi_1),\ldots,(P,\varphi_I)$ defined by $P$ is equal to
$\sup_{n\in\N} T_P^n(\emptyset,\ldots,\emptyset)$ and, because $D$ has $n$
objects only, this least upper bound is obtained after at most $n\times I$
steps. Same proof if all IDBs are tagged (computed as greatest fixed points).

The case where $P$ has $p$ strata is similar: since the IDBs are computed in the order of the strata, assuming stratum $j$ has $I_j$ IDBs, the queries it
 defines will be computed in time $n\times I_j$, hence for the whole of $P$ the complexity will be $n\times \sum_j I_j= n\times I$. The space complexity is clear too
because we have at any time at most $I$ IDBs true of at most $n$ data
objects.

This bound is tight as shown in the next \ref{exstrat}.
\qed

Theorem \refno{strat} subsumes a result of [\cit{gk}],
where it is shown that the data complexity of Monadic Datalog is linear-time;
we prove that 
both the data complexity and the program 
complexity of stratified Monadic inf-Datalog are  linear-time and linear-space:
hence adding greatest fixed points in a stratified way  increases the
expressive power of Monadic Datalog without increasing its evaluation
complexity.
As a consequence of \ref{strat} we get a new unified proof 
of the following result.

\corol The model checking problem for $CTL$, $ETL$ and alternation-free 
Modal $\mu$-calculus can be solved in  time and space $O(|M|\times|f|)$, where
$|M|$ (resp. $|f|$) is the size of the model (resp. the formula);
 hence both the data and program complexities are linear in time and space.
\fin

\preuve Indeed we give in [\cit{gfaa}] a  translation from $CTL$,
 $ETL$ and alternation-free 
Modal $\mu$-calculus into Monadic stratified modal inf-Datalog such that the number of IDBs in the program is less than the size of the formula.
\qed

A unified proof of the time-linearity wrt. the size of the model
  is also given in [\cit{ggv}],
and other proofs are given in [\cit{ces},\cit{cs}].

\figs[struc3]  scaled 600 {A data structure of size 3}

{\petit
\ex[exstrat] Consider the structure given in \ref{struc3}, where
$suc(1,2),suc(2,3),p(1), p(2),\allowbreak  q(3),r(1)$ hold  and the Monadic
Datalog program:
 $$P\colon\left\{\eqalign
           {\varphi (x)&\longleftarrow q(x)\cr
            \varphi (x)&\longleftarrow p(x),suc(x,y), \varphi(x)\cr
              \psi (x)&\longleftarrow \varphi (x),r(x)\cr
              \psi (y)&\longleftarrow \psi (x),suc(x,y)\cr}                    
                    \right. $$
Then, we need 6 steps to compute the queries defined by the program:

$\varphi_0=\emptyset, \varphi_1=\{3\}, \varphi_2=\{2,3\},\varphi_3=\{1,2,3\}
=\varphi_4=\varphi_5=\varphi_6$

$\psi_0=\psi_1=\psi_2=\psi_3=\emptyset, \psi_4=\{1\}, \psi_5=\{1,2\},
\psi_6=\{1,2,3\}$.
\fin

}

We now turn to Monadic inf-Datalog programs with alternations.

\thm[ppal] Let $P$ be a program with $k-1$ alternations of  least fixed
points  and greatest fixed points ($k$ fixed). 
Assume $P$ has $I$  mutually recursive IDBs.
Let $D$ be a relational database having $n$ elements. 
Then the set of {\bf all queries} 
 of the form $(P,\varphi)$, where $\varphi$ is an IDB of $P$,
 can be computed in time 
$O\big((n+1)^{k}\times I\big)$ and space $O(n\times I)$. 
\fin

\preuve Program $P$ has $k-1$ alternations of  least fixed
points  and greatest fixed points, which means that there exist
IDBs  $\varphi_1, {\overline{\varphi_2}},\ldots,\varphi_{k-1},\allowbreak
{\overline{\varphi_k}}$, computed in the order: first $\varphi_1$, then 
${\overline{\varphi_2}}$, \dots, and last ${\overline{\varphi_k}}$.
For simplicity, we first assume that  $I=k$, $k$ even,
  then $P=P_k$ has the following form:

$$ P_k\ \left\{\matrix{
 P'_k\ \ \left\{\matrix{
           {\overline{\varphi_k }}(x)&\longleftarrow \cdots\cr
                         &\vdots\cr
            {\overline{\varphi_k }} (x)&\longleftarrow \cdots\cr
              }\right.\qquad\qquad\qquad\qquad \cr
\noalign{\smallskip}
 {P_{k-1}\ \left\{\matrix{
 P'_{k-1}\ \left\{\matrix{
           \varphi_{k-1}(x)&\longleftarrow \cdots\cr
                         &\vdots\cr
            \varphi_{k-1} (x)&\longleftarrow \cdots\cr
              }\right.\qquad\qquad\cr
\noalign{\medskip}
\vdots\cr 
\noalign{\medskip}
{\qquad\qquad P_2\ \left\{\matrix{
     P'_2\ \left\{\matrix{
           {\overline{\varphi_2 }}(x)&\longleftarrow \cdots\cr
                         &\vdots\cr
            {\overline{\varphi_2 }} (x)&\longleftarrow \cdots\cr
              }\right.\cr
\noalign{\smallskip}
     P_1\ \left\{\matrix{
           \varphi_1 (x)&\longleftarrow \cdots\cr
                         &\vdots\cr
            \varphi_1 (x)&\longleftarrow \cdots\cr
              }\right.\cr}         
        \right.}\cr  }\right.}\cr}\right.
$$

The idea of the algorithm is obtained by adapting an algorithm given in 
[\cit{AN01}]
for evaluating boolean $\mu$-calculus formulas and proceeds as follows.
Let $f_1,\ldots,f_k$ be the queries defined by $\varphi_1,\ldots,{\overline{\varphi_k}}$.
In order to compute $f_k$ we must compute $\inf_i T_{P'_k}^i(\top)$, where
 $\top$ 
is true of every element in the data domain, and $f_k$ will be reached after
at most $n$ steps (because the domain has $n$ elements). However,
since $P'_k$ depends on $\varphi_{k-1}$, we must prealably compute
$f_{k-1}[\top/{\overline{\varphi_k}}]$, which denotes $f_{k-1}$ in which $\top$  
has been substituted for the parameter ${\overline{\varphi_k}}$: this implies computing
$\sup_i T_{P'_{k-1}}^i[\top/{\overline{\varphi_k}}](\emptyset)$,
 which is again reached after at most $n$ steps, etc.
The algorithm is described in \ref{algo1}.

\figs[algo1]  scaled 1000 {Algorithm1}

Notice that in the $k-1$ first nested loops the indices have to go from
1 to $n+1$: indeed each individual $f_j$ is computed in at most
$n$ steps, but then
we have to substitute the value just computed for $f_j$ in $f_1,\ldots,f_{j-1}$
whence the need for one more round of iterations. At the end $f_1,\ldots,f_k$
contain the answers to the queries defined by $\varphi_1,\ldots,
{\overline{\varphi_k}}$.
The complexity of the algorithm is
$(n+1)+(n+1)^2+\cdots+(n+1)^{k-1}+n(n+1)^{k-1}$ which is $O((n+1)^k)$.

The generalization to the case when  $P$ 
has $I$ mutually recursive IDBs, $I > k$,  is straightforward: let  the IDBs of $P$ be for instance
$\Phi = \Phi_1\cup {\overline
{\Phi_2 }}\cup \Phi_3\cup\cdots\cup {\overline{\Phi_{k}}}$.
All the IDBs in $\Phi_i$ (resp. ${\overline{\Phi_{j}}}$ are untagged 
(resp. tagged).
 The order and type of evaluation are as follows: first all IDBs of
$\Phi_1$ are computed as least fixed points, then all IDBs of
${\overline{\Phi_2}}$ are computed as greatest fixed points, \dots ,
 and finally  
all IDBs of $\Phi_{k}$ are computed as greatest fixed points. 
Assume $\Phi_i$ has $m_i$ IDBs, for $i=1,\ldots,k$.

Then it suffices to substitute for instruction:
$f_{{i}}$ := $T_{P'_{i}}(f_{1},f_{2},\ldots, 
                                f_{{i}},\ldots,f_{k})$
the set of $m_i$ instructions:
$$\eqalign{
f_{{i},1} &:= T_{P'_{i},1}(f_{1},f_{2},\ldots, 
                                f_{{i}},\ldots,f_{I})\cr
&\quad\vdots\cr
f_{{i},m_i} &:= T_{P'_{i},m_i}(f_{1},f_{2},\ldots, 
                                f_{{i}},\ldots,f_{I})\cr}$$

where $T_{P'_{i},l}(f_{1},f_{2},\ldots, 
                                f_{{i}-1},\ldots,f_{I})$ denotes the set
of immediate consequences which can be deduced using the rules of $P'_i$ with
head $\varphi_{i,l}$. Now the complexity of the algorithm becomes:

$(n+1)\times m_k+(n+1)^2\times m_{k-1}+\cdots+(n+1)^{k-1}\times m_2
+n(n+1)^{k-1}\times m_1$ which is an $O\big((n+1)^{k}\times \max\{m_i/
i=1,\ldots,k\}\big)\leq O((n+1)^{k}\times I)$.
\qed

We can restate \ref{ppal} as: algorithm1 computes the answers to queries
defined by Monadic inf-Datalog programs with linear-space (data and program)
 complexities, linear-time program  complexity, 
and polynomial-time  data complexity.

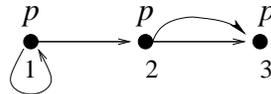
\figs[struc2]  scaled 600 {Another structure of size 3}

{\petit
\ex[exalt] Consider the structure given in \ref{struc2}, where
$suc_1(1,1),suc_0(1,2),suc_0(2,3),\allowbreak  p(1), p(2),p(3)$ hold 
 and the program $P$ below (where $I=k=2$):
 $$P\colon\left\{\eqalign{
  {\overline {\varphi^2} }(x)&\longleftarrow  \theta^1 (x),
                            Suc_0(x,y), Suc_1(x,z),{\overline {\varphi^2}}(y),
                    {\overline {\varphi^2}} (z)\cr
           \theta^1 (x)&\longleftarrow Suc_i(x,y), {\theta^1} (y)\qquad
                            \hbox{ for }i=0,1\cr 
 \theta^1 (x)&\longleftarrow p(x),Suc_i(x,y),{\overline {\varphi^2}} (y)\qquad
                            \hbox{ for }i=0,1\cr
                  } \right. $$
Then algorithm1 will compute: 1. for $f_2=\top$, $f_1= \emptyset, f_1=\{1,2\}$,
and $f_2=\{1,2\}$; then, 2. for $f_2=\{1,2\}$, $f_1= \emptyset, f_1=\{1\}$, and
$f_2=\{1\}$; then, 3. for $f_2=\{1\}$, $f_1= \emptyset, f_1=\{1\}$, and
$f_2=\emptyset$; a last round will give 4.  for $f_2=\emptyset$,
 $f_1=\emptyset$. ($P$  is the translation of the temporal logic formula: 
$\varphi={\bf E}(F^\infty p\wedge {\bf A}\circ\;F^\infty p)$ expressing
  that there exists a path on which $p$ holds infinitely often and moreover,
on all successors of the first state of that path,
 again $p$ holds infinitely often.)
\fin

}

\corol 1. The set of queries defined by Monadic inf-Datalog
programs 
can be computed in time polynomial in the size of the data structure,
 exponential in   the number of alternations of  least fixed
points  and greatest fixed points, and linear in the number of IDBs.
The space complexity is linear in $n\times I$ ($n$ is the size of the
structure and $I$ the number of IDBs).

 2. The model checking problem for the 
Modal $\mu$-calculus can be solved in  time polynomial in  the size of the
model and exponential in the number 
of syntactic alternations of the formula. 
The space complexity is linear in $|M|\times |f|$ ($|M|$ is the size of the
model and $|f|$ the size of the formula).
\fin

\preuve  2 follows from the fact that in [\cit{gfaa}] we gave a 
 translation from  modal $\mu$-calculus 
formulas  into  Monadic (in fact modal) inf-Datalog
programs, such that the number $k$ 
of alternations in the program is equal to the
number of syntactic alternations [\cit{bra99}] of the formula and the number
$I$ of IDBs is less than the size of the formula.
1 is a restatement of \ref{ppal}.
\qed

\section Conclusion

We gave a (linear-) polynomial-time algorithm computing the answers
 to the queries defined by a (stratified) Monadic inf-Datalog program.
The time complexity of this algorithm is $O(n^{k+1})$ where $n$ is the size of the database and $k$ the number of alternations. We believe that this
bound could be slightly improved: indeed there are algorithms
for model checking formulas of  $L\mu_ k$ (which is
equivalent to a fragment of Monadic inf-Datalog), 
with  upper bounds $O\big((n\times |f|)^{k}\big)$ [\cit{EL},\cit{E}] and  
$O\big((n\times |f|)^{2+k/2}\big)$ [\cit{bcjlm}]
(compared to our bound $O\big((n+1)^{k+1}\times |f|\big)$ );
 however the space complexity of the improved
algorithm in [\cit{bcjlm}] becomes exponential whilst the space complexity of the naive algorithms is polynomial [\cit{E}].

\parskip=0pt

\section References

\bibi[AN01]
A.~Arnold, D.~Niwi\'nski,
 {\sl Rudiments of $\mu$-calculus}, Elsevier Science,
Studies in Logic and the Foundations of Mathematics, 146, North-Holland, 
Amsterdam, 2001.

\bibi[bra99] 
J. Bradfield, {\sl
Fixpoint alternation: Arithmetic, transition systems, and the binary tree}, 
 in Theoretical Informatics and Applications, Vol 33,  1999,  341-356.

\bibi[bcjlm]
A. Browne, E. Clarke, S. Jha, D. Long, W. Marrero, {\sl
An improved algorithm for
the evaluation of fixpoint expressions}, {\bf TCS 178} , 1997, 237-255.

\bibi[ces]
E. M. Clarke, E. A. Emerson, A. P. Sistla,  {\sl Automatic
Verification of finite-state concurrent systems using temporal logic
specifications}, ACM TOPLAS, 8,  1986, 244-263.

\bibi[cs]
R. Cleavekand, B. Steffan, {\sl A linear time model checking algorithm for the
alternation-free modal mu-calculus}, Formal methods in system design, 2 (1993),
121-148.

\bibi[e90]
E. Emerson,  {\sl Temporal and modal logic}, Handbook of
Theoretical Computer Science, 1990, 997-1072.
\bibi[E]
E.  Emerson,  {\sl Model Checking and the Mu-Calculus}, in Descriptive
Complexity and Finite Models, N. Immerman and Ph. Kolaitis eds., 
American Mathematical Society, 1997.

\bibi[EL]
E. A. Emerson, C.L.Lei, {\sl
Efficient model checking in fragments
of the propositional $\mu$-calculus}, In Proc. of 1rst Symposium on Logic in
Computer Science, 1986, 267-278.

\bibi[ggv]
G. Gottlob, E. Gr\"adel, H. Veith, {\sl Datalog LITE:
temporal versus deductive reasoning in verification},  
 ACM Trans. on Comput. Logic, 3, 2002, 39-74.

\bibi[gk] G. Gottlob and C. Koch, {\sl 
Monadic Datalog and the Expressive Power of Web Information 
Extraction Languages}, Proc. PODS'02, 17-28.

\bibi[gfaa] I. Guessarian, E. Foustoucos, T. Andronikos, F. Afrati, {\sl
 On Temporal Logic versus Datalog},  to appear in {\bf TCS}, available from
http://www.liafa.jussieu.fr/{\~{ }}ig/gfaa.ps

\bye

\section Appendix

\subsection Datalog

\defi
    A {Datalog program} $\Pi$ is a finite set of function-free Horn
    clauses, called rules, of the form:
$
    \varphi (x_1,\ldots,x_n)\longleftarrow \psi_1(y_{1,1},\ldots,y_{1,n_1}),
    \ldots,\allowbreak \psi_k(y_{k,1},\ldots,y_{k,n_k})
$
    where:

    \item{\bulet}   $x_1,\ldots,x_n$  are variables,
    \item{\bulet}   $y_{i,j}$'s  are either variables or constants,
    \item {\bulet}  $\varphi$ is a predicate, called the  head of the rule, and
    \item {\bulet}   $\psi_1(y_{1,1},\ldots,y_{1,n_1}),
            \ldots,\psi_k(y_{k,1},\ldots,y_{k,n_k})$ is the body of the
            rule,
    \item{\bulet}    A predicate which appears on the head of some rule is called
           intensional database predicate (IDB predicate), whereas the rest
            are called extensional database predicates (EDB predicates),
   \item{\bulet}   $\psi_i$'s  are literals which are either  of the form
            $\varphi_i(y_{i,1},\ldots,y_{i,n_i})$ or of the form
            $\neg \varphi_i(y_{i,1}, \ldots,\allowbreak y_{i,n_i})$, 
           with $\varphi_i$ an EDB predicate, or an IDB predicate; if
           $\psi_i$ is of the form
            $\neg \varphi_i(y_{i,1}, \ldots,\allowbreak y_{i,n_i})$, 
           then  all rules with head $\varphi_i$ have only EDB predicates 
          in their bodies.

The dependency graph of a Datalog program is a directed graph with
nodes the set of IDB predicates of the program; there is an arc from
predicate $\varphi$ to predicate $\psi$ if there is  a rule with head
an instance of $\varphi$ and at least one occurrence of $\psi$ in its
body. Two predicates that belong to the same strongly connected component of
the graph are said to be {\sl mutually recursive}.

\fin

\remarque    We should use the notation Datalog$^{\neg}$ 
for our language, because some 
negations may occur in  rule bodies, but for simplicity sake, 
we will use only Datalog. Notice that  we use negation in a very restrictive 
way (which reduces to
 using negation on EDB predicates only); 
our  Datalog programs are always stratified
with respect to negation, with at most two strata.\fin

    A Datalog program $\Pi$ is said to be {\sl monadic} if all the
predicates occurring in the heads of the rules have arity at most one. 

    A {\sl Datalog query} is  a pair $(\Pi,\varphi)$ consisting 
of a Datalog program $\Pi$ together
with one of its IDB predicates $\varphi$ called {\sl goal} predicate. 

    For any database $D$, $T_{\Pi}(D)$ is the set of ground facts which can be deduced from $D$ by one single application of rules in $\Pi$, and
$\varphi_{\Pi}(D)$ is the set of ground facts about
$\varphi$ which can be deduced from $D$ by applications of the rules in $\Pi$.
The usual semantics of Datalog computes $\varphi_{\Pi}(D)$ by computing 
 least fixed points only. Here, the semantics will be slightly
 different, because some IDB predicates will be ``tagged'' by an
 overline indicating that they must be computed with a greatest fixed
 point ${\overline{\varphi_{\Pi}}}(D)$
instead of a least fixed point $\varphi_{\Pi}(D)$.

\subsection Proof of  \ref{ppal} 

 We prove by induction on $k$ the following slightly
more general lemma: 

\lemme Let $P$ be a program with $k-1$ alternations of  least fixed
points  and greatest fixed points, with IDBs $\varphi_1,
{\overline{\varphi_2}},\ldots,{\overline{\varphi_k}}$,
 and with parameters $g_1,\ldots,g_p$.
($k$ is assumed to be even and the first IDB is assumed to be a least fixed point, but the other cases would be similar.) For any given values of the parameters $g_1,\ldots,g_p$, algorithm1 computes the answers $f_1,\ldots,f_k$ to
the queries  $\varphi_1,
{\overline{\varphi_2}},\ldots,{\overline{\varphi_k}}
$ in time at most $k(n+1)^k$.
\fin

\preuve For $k=1$, the lemma is clear.

Assume it holds for $k-1$ and prove it for $k$. Let $P_{k+1}(g_1,\ldots,g_p)$ 
be the program
$$P_{k+1}(g_1,\ldots,g_p)\ \left\{\matrix
 {P'_{k+1}\ \left\{\eqalign
          {\varphi_{k+1} (x)&\longleftarrow \cdots\cr
                        &\quad\vdots\cr
           \varphi_{k+1} (x)&\longleftarrow \cdots\cr}\right.\cr
\noalign{\smallskip}
            P_{k}(g_1,\ldots,g_p,\varphi_{k+1})\qquad\cr
              }\right.$$
Let $f_1(\varphi_{k+1}),\ldots,f_k(\varphi_{k+1})$ be the answers to the queries defined by $P_{k}(g_1,\ldots,g_p,\varphi_{k+1})$ (with parameters
$g_1,\ldots,g_p$ {\sl and} $\varphi_{k+1}$). Let $P'_{k+1}$ be the rules of 
$P_{k+1}(g_1,\ldots,g_p)$ with head $\varphi_{k+1}$. By definition of the nested fixed points, the answer $f_{k+1}$ to the query defined by $\varphi_{k+1}$ is
the least upper bound of the sequence defined by $f_{k+1}^0=\emptyset$,
$f_{k+1}^1=T_{P'_{k+1}}(f_{k+1}^0)$, \dots , 
$f_{k+1}^n=T_{P'_{k+1}}(f_{k+1}^{n-1})$. This sequence is computed in
 the outermost {\tt FOR} loop of algorithm1 (for $k+1$): indeed, 
$T_{P'_{k+1}}(f_{k+1}^0)=T_{P'_{k+1}}(\emptyset)=
T_{P'_{k+1}}[\emptyset/\varphi_{k+1},f_k(\emptyset)/{\overline{\varphi_k}},
\dots,f_1(\emptyset)/\varphi_1]$, and by the induction hypothesis, the latter
term is computed in time at most $1+k(n+1)^k$ by algorithm1 (for $k$).
Similarly, for $j=1,\ldots,n-1$, each of 
$f_{k+1}^{j+1}=T_{P'_{k+1}}[f_{k+1}^j/\varphi_{k+1},
f_k(f_{k+1}^j)/{\overline{\varphi_k}},
\dots,\allowbreak f_1(f_{k+1}^j)/\varphi_1]$ is computed in time at most $1+k(n+1)^k$.
A final round of iterations is needed to compute
$f_k[f_{k+1}^n/\varphi_{k+1}],
\dots,f_1[f_{k+1}^n/\varphi_{k+1}]$, hence a total time 
$(n+1)(1+k(n+1)^k)< (k+1)(n+1)^{k+1}$.           
\qed

\bye